\documentclass[12pt]{article}
\usepackage{amsmath}
\usepackage{graphicx}
\usepackage{float}
\newcommand{\be}{\begin{equation}}
\newcommand{\ee}{\end{equation}}
\newcommand{\ba}{\begin{eqnarray}}
\newcommand{\ea}{\end{eqnarray}}
\newcommand{\n}[1]{\label{#1}}

\begin{document}
\title{{\bf Hawking Radiation Energy and Entropy\\from a\\Bianchi-Smerlak Semiclassical Black Hole}\!\!
\thanks{Alberta-Thy-11-15, arXiv:1506.01018 [hep-th]}}

\author{Shohreh Abdolrahimi\!\!
\thanks{Internet address: abdolrah@ualberta.ca}
 \, and 
Don N. Page\!\!
\thanks{Internet address:
profdonpage@gmail.com}
\\
Theoretical Physics Institute\\
Department of Physics\\
4-183 CCIS\\
University of Alberta\\
Edmonton, Alberta T6G 2E1\\
Canada
}

\date{2015 September 18}

\maketitle
\large
\begin{abstract}
\baselineskip 20 pt

Eugenio Bianchi and Matteo Smerlak have found a relationship between the Hawking radiation energy and von Neumann entropy in a conformal field emitted by a semiclassical two-dimensional black hole.  We compare this relationship with what might be expected for unitary evolution of a quantum black hole in four and higher dimensions.  If one neglects the expected increase in the radiation entropy over the decrease in the black hole Bekenstein-Hawking $A/4$ entropy that arises from the scattering of the radiation by the barrier near the black hole, the relation works very well, except near the peak of the radiation von Neumann entropy and near the final evaporation.  These discrepancies are calculated and discussed as tiny differences between a semiclassical treatment and a quantum gravity treatment. 

\end{abstract}
\normalsize
\baselineskip 17 pt
\newpage

\section{Introduction}

Eugenio Bianchi and Matteo Smerlak \cite{Bianchi:2014qua, Bianchi:2014vea} have found a beautiful formula relating the Hawking radiation energy flux $F$ and the retarded time derivatives of the von Neumann entanglement entropy $S$ at future null infinity for a two-dimensional conformal field theory in a fixed two-dimensional classical or semiclassical spacetime:
\be
F = \frac{1}{2\pi}\left(\frac{6}{c}\dot{S}^2 + \ddot{S}\right). \n{BS}
\ee
Here $F$ is the energy flux at future null infinity ($\mathcal{I}^{+}$) as a function of the retarded time $u$, $S$ is the renormalized entanglement entropy of the radiation at $\mathcal{I}^{+}$ up to the time $u$, $c$ is a constant that depends on the conformal field, and an overdot represents a derivative with respect to the time $u$.  This formula has been applied to many solvable models of gravitational collapse by Bianchi, De Lorenzo, and Smerlak \cite{Bianchi:2014bma}.

Here we wish to compare the predictions of this formula with what is expected to be the case for the energy flux and the von Neumann entropy of Hawking radiation of massless fields from a four-dimensional spherically symmetric black hole.  For the emission of a conformally invariant scalar field, one might expect the Hawking radiation to be dominated by the scalar field modes that have zero angular momentum (S-waves) and hence are spherically symmetric, effectively reducing the problem to a two-dimensional one for which one might expect the formula of Bianchi and Smerlak to apply, at least to some level of approximation.

We examine whether this is indeed the case.  We find that during most of the Hawking emission by a black hole of initial mass $M_0$ that is large in Planck units, the first term on the right hand side of Eq.\ (\ref{BS}) dominates over the second term, and the power-law dependence of this first term on the mass $M$ is the same as that of the flux $F$ on the left hand side.  Therefore, by choosing the constant $c$ appropriately, one can get a good match between the left and right hand sides.

However, there are some caveats for this description.  

First, for a black hole that starts in a pure state and evaporates away completely by a unitary process that does not lose information, the von Neumann entropy $S$ of the radiation starts at zero at the beginning of the evaporation (when there is not yet any radiation, so that its entropy is zero) and goes back to zero at the end of the evaporation (when all of the information is in the radiation, so that it is in a pure state again with zero von Neumann entropy).  In between, the radiation is entangled with the black hole, so that its von Neumann entropy $S(u)$ rises from zero to a maximum (at what has been called the "Page time" and which we shall denote by $u = u_p$, at which retarded time we shall say the entropy is $S(u_p) = S_p$ and the black hole horizon radius is $R(u_p) = R_p$) and then drops back down to zero.

Because the scattering of modes by the potential barrier near the black hole horizon makes the evaporation a nonadiabatic process, during the stage in which $S(u)$ is rising, its increase is greater, by a factor $b \ge 1$ that is a constant for the massless fields that we shall consider here, than the decrease in the Bekenstein-Hawking entropy $S_{\mathrm {BH}} = A/(4G)$, where $A$ is the area of the black hole event horizon.  (For example, for a four-dimensional Schwarzschild black hole emitting into empty space, numerical calculations \cite{Page:2013dx} have shown that for the emission of just massless photons and gravitons, $b \approx 1.4847$.)  The Bekenstein-Hawking entropy is believed to give the leading term for the logarithm of the number of quantum states of a black hole of area up to $A$ and hence a good approximation for the {\it maximum} von Neumann entropy a black hole of area $A$ can have.  Once the effective number of accessible quantum states for the Hawking radiation exceeds that of the black hole (after $u_p$), the von Neumann entropy of the radiation is limited by $S_{\mathrm {BH}}$ and is expected to be very close to this limit \cite{Page:1993df, Page:1993wv}.  Therefore, before $u_p$, one expects that $\dot{S} \approx -b\dot{S}_{\mathrm {BH}}$, whereas after $u_p$, one expects that $\dot{S} \approx +\dot{S}_{\mathrm {BH}}$.  On the other hand, the energy flux $F$ is proportional to $\dot{S}_{\mathrm {BH}}^2$, with the same coefficient, for both time periods.  Then for Eq.\ (\ref{BS}) to be valid for both periods, $c$ would have to decrease by a factor of $1/b^2$ when the retarded time $u$ crossed $u_p$, which is not consistent with the assumption that $c$ is a constant depending only on the massless fields being emitted.

Second, when $S(u)$ goes from increasing before $u_p$ to decreasing after $u_p$, there is a brief period of retarded time $u$ during which the second term (with the second time derivative of the von Neumann entropy, $\ddot{S}$) on the right hand side of Eq.\ (\ref{BS}), which is negative, dominates over the positive semi-definite first term (with the square of the first time derivative of the von Neumann entropy, $\dot{S}^2$), leading to a negative expression for the energy flux $F$.  If one trusted Eq.\ (\ref{BS}) during this stage of the evaporation, it would seem that the black hole would have to gain a bit of energy at this time, rather than continuously losing energy.

Third, if one extrapolates the semiclassical approximation for $\dot{S} \approx +\dot{S}_{\mathrm {BH}}$ as a function of the black hole mass $M$ down to Planck and sub-Planck values (where the semiclassical approximation is not believed to be valid), one again gets a regime in which the second term on the right hand side of Eq.\ (\ref{BS}) dominates over the first term and again gives a negative expression for the energy flux.

Here we shall discuss these caveats that challenge the validity of Eq.\ (\ref{BS}) for a quantum four-dimensional black hole.  The first caveat is perhaps the most serious, but if it can be swept under the rug, then the second caveat just gives a very small violation of monotonic mass loss and might be an artifact of the approximation in Eq.\ (\ref{BS}) of a single definite (semiclassical) metric.  The third caveat of another violation of monotonic mass loss near the end of the Hawking evaporation can be avoided even more simply by a slight modification of the time dependence of the von Neumann entropy near the final decay of the then-tiny black hole.

One might object (as an anonymous referee has) that there is {\it{a priori}} no reason to expect two-dimensional gravity models to match higher-dimensional gravity quantitatively, so it should not be surprising when they do not.

Indeed (to go beyond the brief objection of the referee), one might give a simple reason why something like Eq.\ (\ref{BS}) holds approximately in $d=2$ spacetime dimensions but not in higher dimensions.  The first term on the right hand side of Eq.\ (\ref{BS}) has an extra power of the entropy $S$ than the second term, whereas both terms have the same number of time derivatives (two).  Therefore, when the entropy is large and when the square of the first time derivative of the logarithm of the entropy is not much smaller than the second time derivative of this logarithm, one would expect the first term to dominate, and then Eq.\ (\ref{BS}) says that the ratio between the energy flux $F$ (the power or luminosity) and the square of the entropy flux $\dot{S}$ should be a constant independent of the gravitational system.

For objects of temperature $T$ and fixed shape with linear size $R$ that is much larger than a thermal wavelength $d$ spacetime dimensions, massless radiation gives an energy flux that goes as $F \sim R^{d-2}T^d$, whereas the entropy flux goes as $\dot{S} \sim R^{d-2}T^{d-1}$.  Therefore, $\dot{S}^2/F \sim (RT)^{d-2}$.  In $d=2$ the right hand side is a constant independent of the size $R$ and the temperature $T$ of the object, so one would indeed expect that the energy flux would be roughly proportional to the square of the entropy flux with a constant of proportionality depending on the number and types of fields in the thermal radiation.

Now for black holes of fixed shape (e.g., spherical, or else rotating with one or more fixed dimensionless rotation parameter that determine the shape), the only quantity that sets the scale is the size $R$, which determines the temperature $T$ to be some size-independent constant divided by $R$, so for asymptotically flat black holes $RT$ is just a shape-dependent constant.  Therefore, in this special case $\dot{S}^2/F \sim (RT)^{d-2}$ for massless radiation is indeed independent of the size of the black hole.

However, if another relevant scale is present, then $RT$ is not necessarily independent of the size of the black hole, and so $\dot{S}^2/F$ can depend on the size.  For example, for spherical black holes in asymptotically anti-de Sitter spacetime that has an additional scale $\ell$ given by the cosmological constant, $RT$ depends on $R/\ell$, and for $R \gg \ell$, $\dot{S}^2/F \sim (R/\ell)^{d-2}$ \cite{Page:2015rxa}, which is not independent of the hole size $R$.

Despite these objections to applying the two-dimensional Bianchi-Smerlak Eq.\ (\ref{BS}) to higher-dimensional black holes, it may be of interest to examine the one idealized case in which it might be hoped to apply, spherical black holes in asymptotically flat spacetime so that the only length scale is that of the black hole and $RT \sim 1$, independent of the one size parameter for such black holes.

\section{Hawking Radiation Energy and Entropy}

We shall generally use units in which the speed of light $c$, the reduced Planck constant $\hbar = h/(2\pi)$, and the Boltzmann constant $k_B$ are set to unity, $c = \hbar = k_B = 1$, but not the Newtonian constant of gravitation $G 
= 6.67384(80)\times 10^{-11}{\mathrm {m}}^3{\mathrm {kg}}^{-1}{\mathrm {s}}^{-2} 
= 2.61210(31)\times 10^{-70}{\mathrm {m}}^2
= 2.90635(35)\times 10^{-87}{\mathrm {s}}^2
= 2.11095(25)\times 10^{15}{\mathrm {kg}}^{-2}$.  However, sometimes we shall refer to Planck units, in which one sets $G = 1$ as well as $c = \hbar = k_B = 1$.   

Let us consider a $d$-dimensional, non-rotating, spherical, uncharged Schwarzschild-Tangherlini black hole of horizon radius $R$, with the metric
\be
ds^2=-(1-\frac{\mu}{r^{d-3}})dt^2+(1-\frac{\mu}{r^{d-3}})^{-1} dr^2+r^2 d\omega_{(d-2)}^2,\n{SHWd}
\ee 
where $\mu = R^{d-3}$ and $d\omega_{(d-2)}^2$ is the metric of a $(d-2)$-dimensional unit sphere. The Arnowitt-Deser-Misner (ADM) mass $M$ of the black hole is given by
\be
M = \frac{(d-2)\Omega_{(d-2)}\mu}{16\pi G}
  = \frac{(d-2)\Omega_{(d-2)}R^{d-3}}{16\pi G},
\ee
where 
\be
\Omega_{(d-2)}=\frac{2\pi^{\frac{d-1}{2}}}{\Gamma(\frac{d-1}{2})}
\ee
is the area of a unit $(d-2)$-dimensional sphere. 
The area of the black hole horizon is 
\be
A=\Omega_{(d-2)} R^{d-2},
\ee 
and the Bekenstein-Hawking thermodynamic entropy is
\be
S_{\mathrm {BH}}
= \frac{A}{4G}=\frac{\Omega_{(d-2)}R^{d-2}}{4G}
= \frac{4\pi}{d-2}\left(\frac{16\pi G}
   {(d-2)\Omega_{(d-2)}}\right)^\frac{1}{d-3} 
   M^\frac{d-2}{d-3}.
\ee

Introducing the null retarded/outgoing coordinate $u=(t-r^*)$ and the advanced/ingoing null coordinate $v=(t+r^*)$, where 
\be
dr^*= \frac{ dr}{\sqrt{1-\frac{\mu}{r^{d-3}}}},
\ee
we can rewrite the metric (\ref{SHWd}) in the following form:
\ba
ds^2=-\left(1-\frac{\mu}{r^{d-3}(u,v)}\right)dudv+r^2 d\omega_{(d-2)}^2.
\ea

Let us consider now that this black hole is evaporating by the process of Hawking radiation to purely massless radiation. Then the black hole mass $M$, horizon radius $R$, parameter $\mu = R^{d-3}$, horizon area $A = \Omega_{(d-2)} R^{d-2}$, and Hawking-Bekenstein entropy $S_{\mathrm {BH}} = A/(4G)$ are all functions of the retarded time $u$. For a spherical uncharged black hole large in Planck units ($GM^2 \gg 1$), the Hawking radiation energy flux of the black hole is
\be
F = -\frac{dM}{du}=\frac{a}{R^2}=\frac{a}{\mu^{2/(d-3)}}
  = a \left(\frac{(d-2)\Omega_{(d-2)}}{16\pi GM}\right)^\frac{2}{d-3},\n{F1}
\ee
where $a$ is a dimensionless constant which can be found from the Hawking emission rate of massless particles. Numerical calculations \cite{Page:1976df,Page:1976ki,Page:1977um,Page:1983ug,Page:2004xp} show that for a four-dimensional black hole, the emission power is approximately $3.3538\times 10^{-5}/(GM)^2$ in photons and $0.3836\times 10^{-5}/(GM)^2$ in gravitons, for a total emission of massless radiation energy about $3.7474\times 10^{-5}/(GM)^2$. This gives $a \approx 0.00014990 \sim 1.5\times 10^{-4}$.

A black hole of initial mass horizon radius $R_0$ and initial mass $M_0 =\\ (d-2)\Omega_{(d-2)}R_0^{d-3}/(16\pi G)$ (say at $u = 0$), emitting purely massless radiation and hence having a Hawking energy flux $F = a/R^2$, will then last a lifetime
\be
u_t = \frac{(d-2)(d-3)\Omega_{(d-2)}}{16\pi(d-1)Ga}R_0^{d-1}
    = \frac{d-3}{(d-1)a}\left(\frac{16\pi G}{(d-2)\Omega_{(d-2)}}
        \right)^{\frac{2}{d-3}} M_0^{\frac{d-1}{d-3}}.
\ee

Then the black hole mass $M(u)$ as a function of $u$ is given by
\be
M(u)=M_0\left(1-\frac{u}{u_t}\right)^{\frac{d-3}{d-1}}.
\ee
The semiclassical approximation for the black hole emission gives the time-dependence of the coarse-grained entropy of the black hole and thus of the emitted radiation as
\be
S_{{\mathrm {rad}}}(u) = b[S_{\mathrm {BH}}(0)-S_{\mathrm {BH}}(u)]
= b S_{\mathrm {BH}}(0)\left[1 
   - \left(1 - \frac{u}{u_t}\right)^{\frac{d-2}{d-1}}\right],\n{Srad0}
\ee
where $S_{\mathrm {BH}}(0) = A_0/(4G) = \Omega_{(d-2)}R_0^{d-2}/(4G)$ is the initial Hawking-Bekenstein entropy of the black hole with mass $M_0$, $S_{\mathrm {BH}}(u)$ is the Hawking-Bekenstein entropy of the black hole when its mass is $M(u)$, and $b$ is the ratio by which the increase in
the coarse-grained entropy of the Hawking radiation (e.g., ignoring its
entanglement with the black hole) is greater than the decrease in the
coarse-grained Bekenstein-Hawking entropy of the black hole.  Numerical calculations \cite{Page:1976ki,Page:2013dx} show that for a massive 4-dimensional Schwarzschild black hole that emits essentially only photons and gravitons, $b \approx 1.48472$.

We assume that the semiclassical Bekenstein-Hawking entropy of a nonrotating uncharged black hole is a good approximation for the maximum von Neumann
entropy of a black hole of the same energy, at least if one neglects the entropy
associated with the location and/or motion of the black hole in a
space sufficiently large that this could in principle rival the Bekenstein-
Hawking entropy. Second, the semiclassical entropy
calculated for the Hawking radiation is the maximum von Neumann
entropy for radiation with the same expectation value of the
number of particles in each of the modes.
Under the extra assumptions that the black hole starts in essentially
a pure state, that the Hawking evaporation is a unitary process,
and that we can neglect the interaction with other systems, the von Neumann entropy of the evaporating black hole equals that of the Hawking radiation
that has been emitted. 

The Bekenstein entropy of the black hole decreases monotonically with time and the radiation entropy ($\ref{Srad0}$) increases monotonically with time. The two values cross at the so-called `Page time,' at
\be
u_p = u_t\left[1 - \left(\frac{b}{b+1}\right)^{\frac{d-1}{d-2}}\right].
\ee
Then when one applies the assumption of unitarity to the results of the semiclassical approximation, the von Neumann entropy of the radiation during the entire evaporation is expected to be given to a good approximation by the minimum of the dimensions of the radiation and black hole subspaces of the Hilbert space \cite{Page:2013dx}, so
\ba
S \approx S_{\mathrm {approx}}
  = b[S_{\mathrm {BH}}(0)-S_{\mathrm {BH}}(u)]\theta(u_p - u)
    + S_{\mathrm {BH}}(u)\theta(u - u_p)\nonumber\\
  = b S_{\mathrm {BH}}(0)
           \left[1 - \left(1 - \frac{u}{u_t}\right)^{\frac{d-2}{d-1}}\right]
            \theta(u_p - u)
	 + S_{\mathrm {BH}}(0)
	   \left(1 - \frac{u}{u_t}\right)^{\frac{d-2}{d-1}}
            \theta(u - u_p), \n{Sapprox}
\ea
where $\theta$ is the Heaviside step function.  The maximum of the von Neumann entropy is at $u=u_p$, where the coefficients of the two step functions above coincide.

\section{Comparison with the Bianchi-Smerlak Formula}

Now let us compare the approximations above for the energy flux, Eq.\ (\ref{F1}), and entropy, Eq.\ (\ref{Srad0}), of the Hawking radiation with the Bianchi-Smerlak formula (\ref{BS}).

The Hawking temperature for the $d$-dimensional spherical uncharged Schwarzschild-Tangherlini black hole of horizon radius $R$ is the surface gravity $\kappa$ divided by $2\pi$.  With $-g_{00} = g^{rr} = V = 1 - \mu/r^{d-3} = 1 - R^{d-3}/r^{d-3}$ and with the radial proper length element $dl = \sqrt{g_{rr}}dr$, the surface gravity is the following expression evaluated on the horizon, $r=R$:
\be
\kappa = \frac{d\sqrt{-g_{00}}}{dl} = \frac{d(V^{1/2})}{V^{-1/2}dr}
      = \frac{1}{2}\frac{dV}{dr} = \frac{d-3}{2}\frac{R^{d-3}}{r^{d-2}}.
\ee
This then gives
\be
T = \frac{\kappa}{2\pi} = \frac{d-3}{4\pi R}.
\ee.

Now the first law of black hole thermodynamics gives $dM = TdS_{\mathrm {BH}}$, so $\dot{M} = - F = - a/R^2$ gives
\be
\dot{S}_{\mathrm {BH}}=\frac{\dot{M}}{T} = -\frac{F}{T} = -\frac{4\pi a}{(d-3)R}.
\ee
From $M = (d-2)\Omega_{(d-2)}R^{d-3}/(16\pi G)$, one also gets
\be
\dot{R} = \frac{-16\pi G a}{(d-2)(d-3)\Omega_{(d-2)}R^{d-2}}
\ee
and hence
\be
\ddot{S}_{\mathrm {BH}} = -\frac{64\pi^2 G a^2}{(d-2)(d-3)^2\Omega_{(d-2)}R^d}.
\ee
Then from Eq. (\ref{Sapprox}), one gets
\be
\dot{S} \approx \dot{S}_{\mathrm {approx}}
  = + \frac{4\pi a b}{(d-3)R}\theta(u_p - u)
    - \frac{4\pi a}{(d-3)R}\theta(u - u_p), \n{dSapprox}
\ee
and
\ba
\ddot{S} \approx \ddot{S}_{\mathrm {approx}}
  = + \frac{64\pi^2 G a^2 b}{(d-2)(d-3)^2 \Omega_{(d-2)}R^d}\theta(u_p - u)
  \nonumber\\
    - \frac{64\pi^2 G a^2}{(d-2)(d-3)^2\Omega_{(d-2)}R^d}\theta(u - u_p)
    - \frac{4\pi a (b+1)}{(d-3)R_p}\delta(u - u_p). \n{ddSapprox}
\ea
Plugging this into the Bianchi-Smerlak formula (\ref{BS}) then gives
\ba
F \approx F_{\mathrm {BS}} &=& \frac{1}{2\pi}\left(\frac{6}{c}\dot{S}_{\mathrm {approx}}^2 + \ddot{S}_{\mathrm {approx}}\right)
\nonumber\\
&=& \left(\frac{48\pi a^2 b^2}{c(d-3)^2 R^2}
   +\frac{32\pi G a^2 b}{(d-2)(d-3)^2 \Omega_{(d-2)}R^d}\right)\theta(u_p - u)
  \nonumber\\
&+& \left(\frac{48\pi a^2}{c(d-3)^2 R^2}
      -\frac{32\pi G a^2}{(d-2)(d-3)^2 \Omega_{(d-2)}R^d}\right)\theta(u - u_p)
  \nonumber\\
&-& \frac{2 a (b+1)}{(d-3)R_p}\delta(u - u_p). \n{Fapprox}
\ea

For $u < u_p$, the ratio of the second term (proportional to $\ddot{S}$) to the first term (proportional to $\dot{S}^2$) is $c/[6b(d-2)S_{\mathrm {BH}}]$; for $u > u_p$ it is $-c/[6(d-2)S_{\mathrm {BH}}]$.  Therefore, except very near $u = u_p$ (the so-called Page time when the von Neumann entropy of the radiation reaches its maximum), where Eq. (\ref{Sapprox}) is not a good approximation to the von Neumann entropy $S$ that actually rounds off rather than having a kink (see below), the second term is much smaller than the first term whenever the black hole is large in Planck units, so that $S_{\mathrm {BH}} \gg 1$, which breaks down near $u = u_t$, near the final evaporation of the black hole.  That is, except for the retarded time $u$ near either $u_p$ or $u_t$,
\be
F \approx F_{\mathrm {BS}} \approx \frac{48\pi a^2}{c(d-3)^2 R^2}
   [b^2 \theta(u_p - u) + \theta(u - u_p)].
\ee
This would fit well with what one expects for a large black hole emitting almost entirely massless radiation, $F = a/R^2$ by Eq. (\ref{F1}) if
\be
c = \frac{48\pi a}{(d-3)^2}[b^2 \theta(u_p - u) + \theta(u - u_p)]. \n{c}
\ee

However, $c$ is supposed to be a constant, not changing between the periods $u < u_p$ and $u > u_p$, which is in conflict with Eq. (\ref{c}) if $b\neq 1$.  Therefore, the Bianchi-Smerlak formula (\ref{F1}) does not seem to work well for a black hole evaporating so that during the early stage the von Neumann entropy of the radiation is larger by a factor $b > 1$ than the decrease in the Bekenstein-Hawking entropy $S_{\mathrm {BH}} = A/(4G)$ of the black hole (not the von Neumann entropy during this stage, though it is to a good approximation for $u > u_p$).  This does seem to be a serious problem with applying the Bianchi-Smerlak formula (actually derived just for emission in two dimensions from a fixed spacetime metric) to black holes in $d > 3$ dimensions with a quantum spacetime.

See Figure 1 for a plot of $M(u)$, the time integral of $-F = -a/R^2$ and of $-F_{\mathrm {BS}}$ for the two constant values of $c$, neither of which matches that of the expected $-F = -a/R^2$.

\begin{figure}[H]
\centering
\includegraphics[width=1\textwidth]{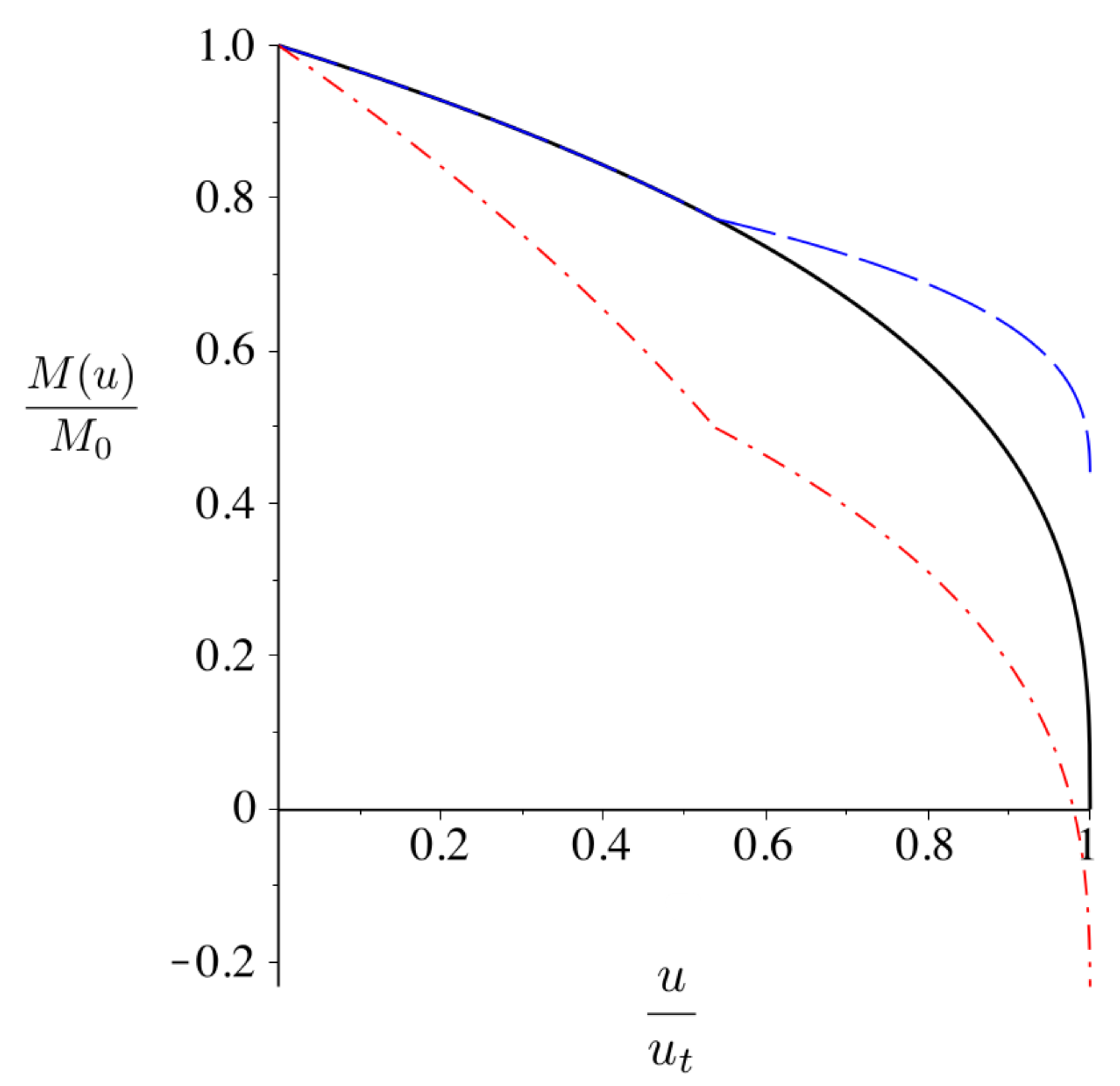}
\caption{Plot of black hole mass vs.\ time in 4-dimensions.
Solid black line is the expected $M(u)$ for $dM/du = - F = - a/(2M)^2$.
Dashed blue line is $M(u)$ for $dM/du = - F_{\mathrm {BS}}$ with $c = 48\pi a$.
Dashed-dotted red line is $M(u)$ for $dM/du = - F_{\mathrm {BS}}$ with $c = 48\pi a b^2.$}
\end{figure}

Nevertheless, let us proceed by sweeping this problem under the rug and proceeding to the next issue, which is how the Bianchi-Smerlak formula for the flux behaves near the peak in $S_{\mathrm {approx}}$, which is at $u = u_p$.  If we use Eq. (\ref{Sapprox}) for the entropy $S$, this has a kink in it (even if we set $b=1$) at $u = u_p$, so the resulting approximation (\ref{Fapprox}) for the flux $F$ has a negative delta function at $u = u_p$, giving a sudden increase in the black hole mass.  In reality $S(u)$ will be smooth near $u = u_p$, so let us look for an improved approximation giving such a smooth $S(u)$.

In \cite{Page:1993df,Page:1993wv}, it was noted that if one has a random pure state (using the Haar measure) in a Hilbert space that is the tensor product of two subsystem Hilbert spaces of dimensions $m$ and $n$, both large, then for $m \leq n$ the average von Neumann entropy of each subsystem is
\be
S \approx \ln{m} - \frac{m}{2n}. \n{subsystem}
\ee

For a black hole that is initially in a pure state surrounded by vacuum (another pure state), we can take $m$ and $n$ to be the effective Hilbert space dimensions of the black hole (using the exponential of the Bekenstein-Hawking entropy $S_{\mathrm {BH}} = A/(4G)$ as an estimate of the dimension of the Hilbert space of black holes of horizon areas close to $A$) and of the radiation (using the exponential of the thermodynamic radiation entropy $S_{\mathrm {rad}} = b[S_{\mathrm {BH}}(0)-S_{\mathrm {BH}}(u)]$ as an estimate of the effective dimension of the Hilbert space of the radiation that would be entangled with the hole; if the black hole is emitting into asymptotically flat empty space, the total Hilbert space dimension for the fields in this space would be infinite, even for bounded total energy, but nearly all of that Hilbert space would develop negligible entanglement with the black hole and hence can be assumed to be in nearly a pure vacuum state and not contribute much to the entanglement entropy).  

For brevity in the equations below, let the Bekenstein-Hawking thermodynamic entropy of the black hole be
\be
s \equiv S_{\mathrm {BH}}(u);\ \ s_0 \equiv S_{\mathrm {BH}}(0).
\ee
Then the radiation entropy is approximately $b(s_0 - s)$.

Therefore, for $u < u_p$, the effective dimension of the smaller subsystem (the radiation) is $m \approx \exp{(bs_0-bs)}$, and that of the larger subsystem (the black hole) is $n \approx \exp{(s)}$, whereas for $u > u_p$, $m$ and $n$ are reversed.  Eq. (\ref{subsystem}) then gives
\ba
S \approx S_{\mathrm {smooth}} = 
    \left[bs_0-bs - \frac{1}{2}\exp{(bs_0-bs-s)}\right]\theta(u_p - u)
     \nonumber\\
    +\left[s - \frac{1}{2}\exp{(s+bs-bs_0)}\right]\theta(u - u_p). \n{Ssmooth}
\ea 
It is convenient to define
\be
x \equiv bs_0 - (b+1)s = (b+1)(s_p-s),
\ee
where
\be
s_p = \frac{b}{b+1}s_0
\ee
is the value of $\ln{m} = \ln{n}$ where they are equal, at $u = u_p$.  This gives $x < 0$ for $u < u_p$, and $x > 0$ for $u > u_p$.  Then Eq. (\ref{Ssmooth}) gives the von Neumann entropy as
\be
S \approx \left(s_p + \frac{b}{b+1}x - \frac{1}{2}e^x\right)\theta(-x)
   +\left(s_p - \frac{1}{b+1}x - \frac{1}{2}e^{-x}\right)\theta(x).
\label{4d-entropy}
\ee
The first two time ($u$) derivatives of the von Neumann entropy are
\be
\dot{S} \approx \left(\frac{b}{b+1}-\frac{1}{2}e^x\right)\dot{x}\theta(-x)
        +\left(-\frac{1}{b+1}+\frac{1}{2}e^{-x}\right)\dot{x}\theta(x),
\ee
\be
\ddot{S} \approx \left[-\frac{1}{2}e^x\dot{x}^2
+\left(\frac{b}{b+1}-\frac{1}{2}e^x\right)\ddot{x}\right]\theta(-x)
    +\left[-\frac{1}{2}e^{-x}\dot{x}^2
+\left(-\frac{1}{b+1}+\frac{1}{2}e^{-x}\right)\ddot{x}\right]\theta(x),
\ee
where
\be
\dot{x} = -(b+1)\dot{s} \equiv -(b+1)\frac{d}{du}S_{\mathrm {BH}}(u)
        = \frac{4\pi a(b+1)}{(d-3)R},
\ee
\be
\ddot{x} = -(b+1)\ddot{s} \equiv -(b+1)\frac{d^2}{du^2}S_{\mathrm {BH}}(u)
         = \frac{\dot{x}^2}{(b+1)(d-2)s}.
\ee
One can see that $S$, $\dot{S}$, and $\ddot{S}$ are all continuous at $u = u_p$, but the third derivative, $d^3S/du^3$, is not.  Also, one can see that, except near the final stage of the Hawking radiation, when the area $A$ gets near the Planck value that is the Newtonian gravitational constant $G$ in our natural units with $\hbar = c = k_B = 1$, the Bekenstein-Hawking entropy $s$ is very large, so $|\ddot{x}| \ll \dot{x}^2$, meaning that we can effectively ignore the $\ddot{x}$ terms.

Plugging these equations into the Bianchi-Smerlak Eq. (\ref{BS}) then give the flux
\ba
F \approx \left[\frac{3}{4\pi c}\left(\frac{2b}{b+1}-e^x\right)^2 \dot{x}^2
     -\frac{1}{4\pi}e^x\dot{x}^2
      +\frac{1}{4\pi}\left(\frac{2b}{b+1}-e^x\right)\ddot{x}\right]
      \theta(-x)
      \nonumber\\
    +\left[\frac{3}{4\pi c}\left(\frac{2}{b+1}-e^{-x}\right)^2 \dot{x}^2
     -\frac{1}{4\pi}e^{-x}\dot{x}^2
      +\frac{1}{4\pi}\left(-\frac{2}{b+1}+e^{-x}\right)\ddot{x}\right]
      \theta(x). \n{flux}
\ea
Because of the continuity of $\dot{S}$ and $\ddot{S}$, the flux $F = -\dot{M} \equiv -dM/du$ is also continuous at $u = u_p$, but $\dot{F} \equiv dF/du = - d^2M/du^2$ is not.  If we do drop the $\ddot{x}$ terms in the expression above for $F$, we get
\be
F \approx 
\frac{\dot{x}^2}{4\pi}\left\{\left[\frac{3}{c}\left(\frac{2b}{b+1}-e^x\right)^2
 -e^x
\right]\theta(-x)
+\left[\frac{3}{c}\left(\frac{2}{b+1}-e^{-x}\right)^2-e^{-x}
\right]\theta(x)\right\}. \n{Fmiddle}
\ee

See Figure 2 for a plot of the normalized Bianchi-Smerlak flux $(R^2/a)F_{\mathrm {BS}}$, for $d = 4$ and $c = 48\pi a$, versus $S_{\mathrm {BH}}(u_p)-S_{\mathrm {BH}}(u) = x/(b+1)$.

\begin{figure}[H]
\centering
\includegraphics[width=1\textwidth]{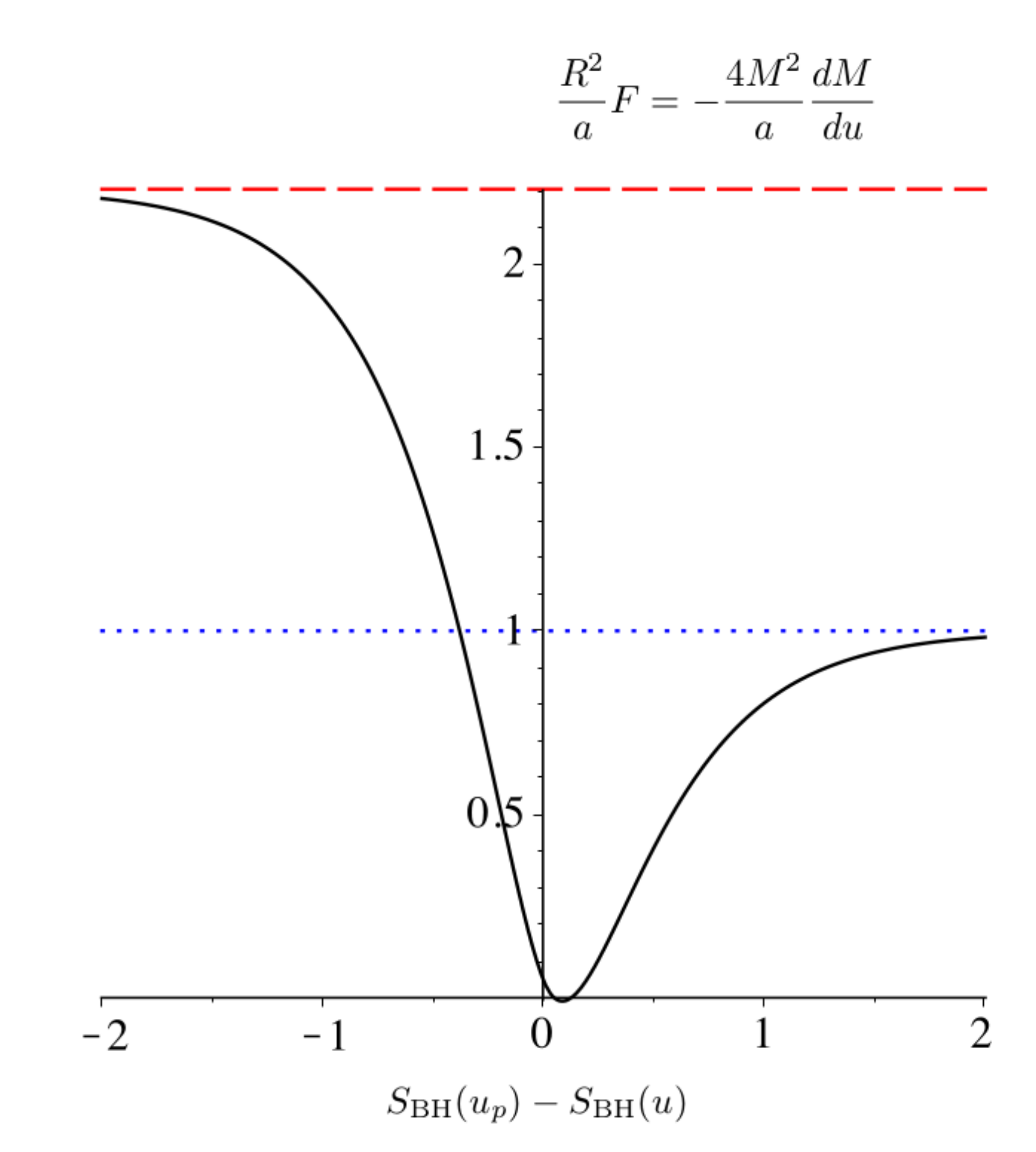}
\caption{Plot of Bianchi-Smerlak flux formula near the peak of $S$.
Solid line is $(R^2/a)F_{\mathrm {BS}}$ for $d = 4$ and $c = 48\pi a$.
Dashed red line is the asymptotic value for $u-u_p \ll -R$, a factor of $b^2$ too large with this constant value of $c$.
Dotted blue line is the asymptotic value for $u-u_p \gg R$, matching the expected value of the normalized flux.}
\end{figure}

\newpage

Since $S(u)$ is turning over from increasing to decreasing, there is always a range in $x = bs_0 - (b+1)s$ for which the $(6/c)\dot{S}^2$ term in the Bianchi-Smerlak Eq. (\ref{BS}) for the flux $F$ is smaller in magnitude than the negative $\ddot{S}$ term, so $F$ is negative in this range.  For $[(b-1)/(b+1)]^2 > c/3$, which is the case for large 4-dimensional black holes emitting just photons and gravitons, the range is entirely in $x > 0$ and is
\be
\ln{\left(\frac{b+1}{2}\right)} - 2\ln{(\sqrt{1+\delta}+\sqrt{\delta})} < x <
\ln{\left(\frac{b+1}{2}\right)} + 2\ln{(\sqrt{1+\delta}+\sqrt{\delta})},
\ee
where
\be
\delta = \frac{(b+1)c}{24} = \frac{2\pi a(b+1)}{(d-3)^2} \approx 0.0023402,
\ee
using Eq. (\ref{c}) for $c = 48\pi a/(d-3)^2$ for $u > u_p$ where the $x > 0$ region with negative $F$ occurs, and then at the end (as shall also be done in successive equations below) inserting the numerical values above for $a$ and for $b$ for a large 4-dimensional Schwarzschild black hole that emits only photons and gravitons.

If we define
\be
y \equiv 4\ln{(\sqrt{1+\delta}+\sqrt{\delta})} = 4\sinh^{-1}{\sqrt{\delta}}
\approx 0.19343,
\ee
then integrating the negative flux over the time period $[(d-3)yR_p]/[4\pi a(b+1)] \approx 41 R_p$ over which it occurs (with $R_p$ being the radius of the black hole horizon then) gives a mass increase during this period of
\be
\Delta M = T_p \left(\frac{\sinh{y} - y}{b+1}\right) \approx 0.0004863 T_p,
\ee
where $T_p = (d-3)/(4\pi R_p)$ is the Hawking temperature of the black hole at the time very near $u = u_p$ when the von Neumann entropy $S$ is maximized and when the flux $F = - dM/du$ is negative, leading to the increase in the mass $M$.

We see that according to the Bianchi-Smerlak formula Eq.\ (\ref{BS}), the mass would indeed increase, but the increase would be extremely small for a large-mass black hole in 4 dimensions, less than 0.05\% of the energy $T_p$ of one thermal photon at the Hawking temperature.

Furthermore, because of random root-$N$ fluctuations in the number of particles emitted, which in 4-dimensions is $\sqrt{N} \sim R/\sqrt{G}$, with each particle taking a time $\sim R$ to be emitted, one might consider the Hawking evaporation to give an ensemble of semiclassical spacetimes, each with its own time at which the von Neumann entropy of the radiation is maximized.  The rms spread in these times would be of the order of $\sim R^2/\sqrt{G}$, which for a solar-mass black hole of about $10^{38}$ Planck masses would be of the order of $10^{76}$ Planck times or $10^{15}$ times the present age of the universe.  If the Bianchi-Smerlak formula applied to each spacetime in the ensemble, then there would be a negative mass flux of energy $\sim -0.0005\, T_p$ over a time less than a millisecond but uncertain by a time of the order of $10^{38}$ times the duration of the negative flux, a time roughly fifteen orders of magnitude larger than the present age of the universe.  Therefore, the effect of this negative energy flux, if it really occurs, seems to be quite negligible and virtually impossible to detect.  Surely it would be washed out by any averaging over the uncertain time at which a random member of the ensemble of radiating black hole spacetimes reached the point where one might expect the von Neumann entropy of the radiation to be maximized.

One can also ask how small one can make the negative energy emitted in a two-dimensional black hole mirror model in which one has control of the motion of the mirror and hence of the time evolution of the entropy, $S(u)$.  In the four-dimensional black hole example above in which $S(u)$ was the specific function given by Eq.\ (\ref{4d-entropy}), we found that the negative energy emitted was just a small fraction (but still of the general order of magnitude of unity) of the energy of a single thermal quantum.  However, by choosing the mirror trajectory appropriately for a two-dimensional model, we can get the negative energy emitted to be of the order of a thermal quantum divided by the maximum von Neumann entropy and hence enormously smaller than one thermal quantum.  For example, consider the following model with four retarded time periods, each of length $\tau$:

(1) For $-2\tau < u < -\tau$, let $S(u) = S_1(u/\tau + 2)^2$, which goes from 0 at $u = - 2\tau$ to $S_1$ at $u = -\tau$.  $\dot{S} = (2S_1/\tau)(u/\tau + 2)$ goes from 0 at $u = - 2\tau$ to $2S_1/\tau$ at $u = -\tau$.  $\ddot{S} = 2S_1/\tau^2$ is a constant during this period.

(2) For $-\tau < u < 0$, let $S(u) = S_1 + (c/6)\ln{[1 + (12 S_1/c)(u/\tau + 1)]}$, which goes from $S_1$ at $u = -\tau$ to $S_1 + (c/6)\ln{(1 + 12S_1/c)}$ at $u = 0$.  $\dot{S} = (2S_1/\tau)/[1 + (12S_1/c)(u/\tau + 1)]$ goes from $2S_1/\tau$ at $u = -\tau$ to $(2S_1/\tau)/(1 + 12S_1/c)$ at $u = 0-$.  $\ddot{S} = - [24(S_1/\tau)^2/c]/[1 + (12S_1/c)(u/\tau + 1)] = -(6/c)\dot{S}^2$.

Now one makes the entropy a symmetric function of the retarded time $u$, so $S(u) = S(-u)$.  At $u = 0$, $\dot{S}$ jumps from $(2S_1/\tau)/(1 + 12S_1/c)$ at $u = 0-$ to $-(2S_1/\tau)/(1 + 12S_1/c)$ at $u = 0+$, so to cover the period $-\tau < u < \tau$ during which $\dot{S}^2$ is continuous, $\ddot{S} = -(6/c)\dot{S}^2 - (4S_1/\tau)/(1 + 12S_1/c)\delta(u)$.  At the other two junctions between the four regions, both $S$ and $\dot{S}$ are continuous functions of $u$, though $\ddot{S}$ is not.

(3) For $0 < u < \tau$, $S(u) = S_1 + (c/6)\ln{[1 - (12 S_1/c)(u/\tau - 1)]}$.

(4) For $\tau < u < 2\tau$, $S(u) = S_1(u/\tau - 2)^2$.

Now the Bianchi-Smerlak Eq. (\ref{BS}) gives the energy flux as
\ba
F = \left[\frac{12S_1^2}{\pi c\tau^2}\left(\frac{|u|-2\tau}{\tau}\right)^2
          +\frac{S_1}{\pi\tau^2}\right]
    \theta(2\tau - |u|)\theta(|u| - \tau)
    -\frac{4 c S_1/\tau}{2\pi(12 S_1 + c)} \delta(u),
\ea
which is positive for $\tau < |u| < 2\tau$, zero for $0 < |u| < \tau$, and has a negative delta function at $u = 0$.  Positive energy $E_1 + E_4 = (8 S_1^2 + 2 c S_1)/(\pi c \tau)$ is emitted during the periods (1) and (4) with $\tau < |u| < 2\tau$, and negative energy $-E_{\rm{negative}} = - 4 c S_1/[2\pi\tau(12 S_1 + c)]$ is emitted at $u=0$ when $\dot{S}$ jumps downward.

When $S_1$ is very large in comparison with 1, $c$, and $c^2$, the maximum von Neumann entropy is $S_{\rm{max}} = S_1 + (c/6)\ln{(1 + 12S_1/c)} \sim S_1$, and the total energy emitted is $E = E_1 + E_4 - E_{\rm{negative}} = (8 S_1^2 + 2 c S_1)/(\pi c \tau) - 4 c S_1/[2\pi\tau(12 S_1 + c)] \sim 8 S_1^2/(\pi c \tau) \sim 8 S_{\rm{max}}^2/(\pi c \tau)$, which is approximately a factor of $16/3$ larger than the minimum energy $E_{\rm{min}} = 6 S_{\rm{max}}^2/(\pi c \Delta u)$ that can be emitted during a period of retarded time $\Delta u$ (here $4\tau$) if $S$ and $\dot{S}$ both start at zero at the beginning of the period and end at zero at the end of the period.  (The minimum is attained by having $S$ increase uniformly from zero to $S_{\rm{max}}$ during the first half of the total retarded time period $\Delta u$ and then decrease uniformly back to zero during the second half of the period.)

However, the magnitude of the negative energy emitted in the example above is only $E_{\rm{negative}} \sim c/(6\pi\tau)$, with no factor of $S_{\rm{max}}$, so it is smaller by a factor of the order of $1/S_{\rm{max}}^2$ than the magnitude of the positive energy emitted.  To put it another way, one might say that an average quantum emitted has energy $\omega \sim E/S_{\rm{max}} \sim 8 S_{\rm{max}}/(\pi c\tau)$, and in comparison, $E_{\rm{negative}}/\omega \sim c^2/(48 S_1) \ll 1$.  That is, in the model above, the negative energy emitted is much smaller than the average energy of a quantum emitted, smaller by a factor of very roughly the reciprocal of the maximum von Neumann entropy.  In this sense the negative energy flux implied by the Bianchi-Smerlak Eq. (\ref{BS}) when the von Neumann entropy $S(u)$ is at or near its maximum value can be made quite negligible if this von Neumann entropy is very large.

Now let us go back to the four-dimensional black hole model and turn to the final stages of its Hawking evaporation.
If one takes the approximate formula Eq.\ (\ref{Sapprox}) for the von Neumann entropy that is valid except near $u = u_p$ and inserts it into the Bianchi-Smerlak formula Eq.\ (\ref{BS}) for the flux to get Eq.\ (\ref{Fapprox}), one finds that one also gets a negative flux near the end at the black hole lifetime, at $u \sim u_t$.  However, one would not expect that Eq.\ (\ref{Sapprox}) would be valid when the black hole has evaporated down to near the Planck size.  One does not know what the actual expressions should be (which also should take into account the uncertainty in the time $u_t$, analogous to the uncertainty in the time $u_p$, that would be of the order of $u_t/\sqrt{S_{\mathrm {BH}}(0)} \sim R_0^{d/2}/\sqrt{G}$, an expression valid in any number of spacetime dimensions greater than 3 where we can have black hole evaporation).  Nevertheless, one can easily come up with expressions that give a positive energy flux all the way down to zero mass and size for the black hole.

For simplicity, we shall just consider $d=4$.  If one defines $X \equiv R^2/(8Ga)$, then for a large Schwarzschild black hole, which has $X \gg 1$, the Bekenstein-Hawking entropy is $S_{\mathrm {BH}} = \pi R^2/G = 8\pi a X$, and the Hawking evaporation gives
\be
\dot{R} = 2G\dot{M} = - \frac{2Ga}{R^2} = - \frac{1}{4X}. \n{Rdot}
\ee
If now one keeps this Eq.\ (\ref{Rdot}) for $\dot{R}$ but replaces the Bekenstein-Hawking entropy formula by
\be
S = \frac{8\pi a X}{1 + 1/X} 
  = \frac{S_{\mathrm {BH}}^2}{S_{\mathrm {BH}}+8\pi a},
\ee
one can calculate that the Bianchi-Smerlak formula Eq.\ (\ref{BS}) with $c = 48\pi a$ gives a flux of the form
\be
F = \frac{a}{R^2}\left[1 - \frac{X^3+8X^2+9X}{2(X+1)^4}\right]. \n{factor}
\ee
This never goes negative but instead increases monotonically with the time $u$ as $R$ decreases, always remaining within a factor of 3 of the expression $F = a/R^2$ from Eq.\ (\ref{F1}).  Therefore, there is no problem constructing $S(u)$ that is close to the Bekenstein-Hawking black hole entropy $S_{\mathrm {BH}} = A/(4G)$ for a large hole but deviates from it when the hole gets small so that the Bianchi-Smerlak formula for the flux stays positive no matter how small the black hole gets.  The flux can even have the asymptotic form $F = a/R^2$ for very small $R^2/G = 8 a X$ as well as for very large $R^2/G$, as the example above shows.

See Figure 3 for a plot of the factor in the square brackets in Eq.\ (\ref{factor}), $(R^2/a)F$, versus the retarded time $u-u_t$, near the end of the black hole evaporation.

\begin{figure}[H]
\centering
\includegraphics[width=1\textwidth]{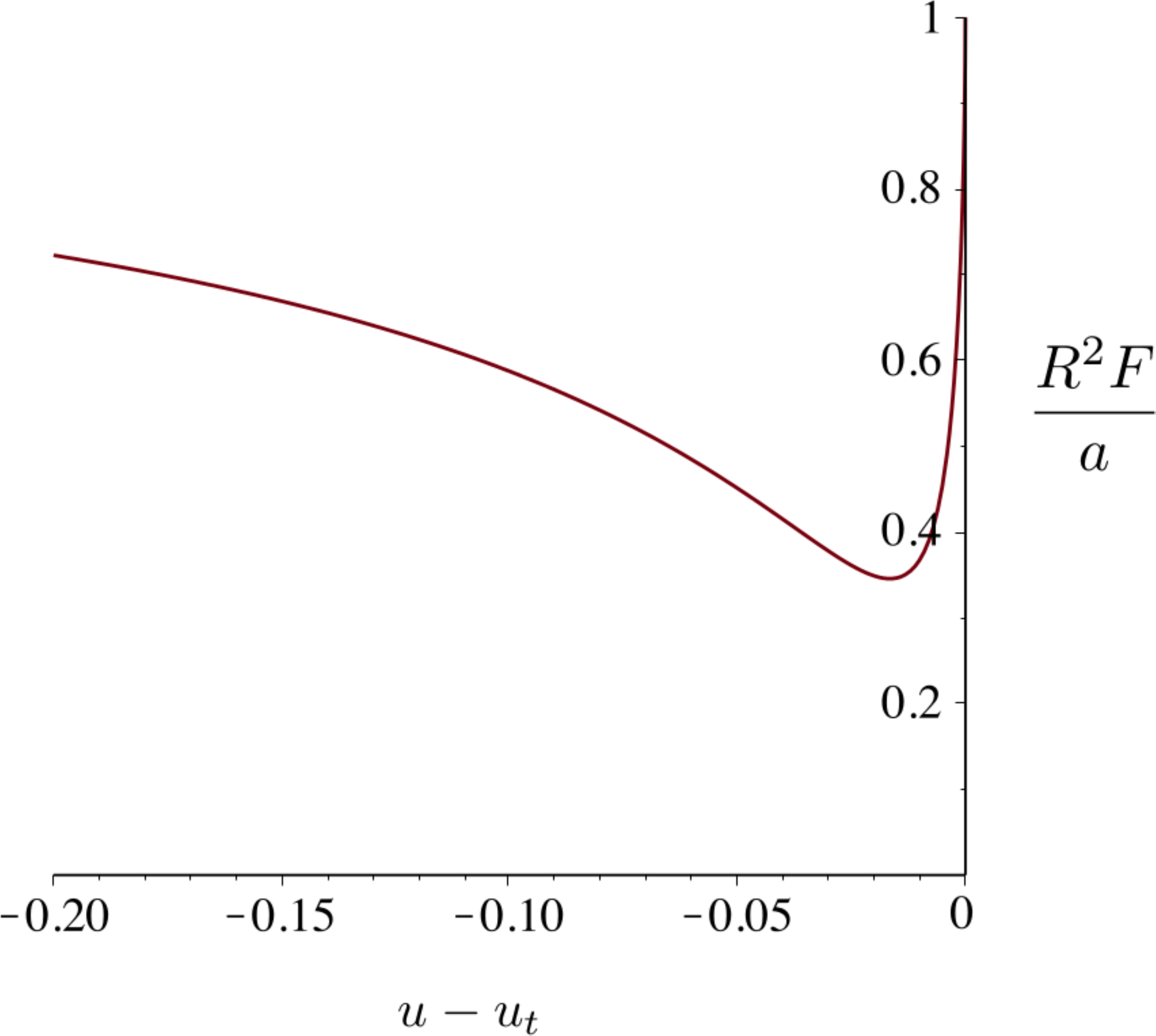}
\caption{Plot of the normalized Bianchi-Smerlak flux near the end of the Hawking radiation:  $(R^2/a)F$ versus retarded time $u-u_t$, for $S = S_{\mathrm {BH}}^2/(S_{\mathrm {BH}}+8\pi a)$ instead of for $S = S_{\mathrm {BH}}$.}
\end{figure}

\newpage

\section{Conclusions}

In conclusion, the Bianchi-Smerlak formula Eq.\ (\ref{BS}), which was derived for a 2-dimensional model black hole with quantum field theory on a definite spacetime metric (say with boundary conditions given by a moving mirror), seems to work fairly well for quantum black holes in higher dimensions, but there are some inadequacies in the model.  Perhaps the most serious is the fact that one cannot match the entropy and energy flux both before and after the peak in the von Neumann entropy with a fixed value of the constant $c$ in the Bianchi-Smerlak formula if the black hole evaporation process is not adiabatic but generates extra entropy in the radiation in excess of the decrease in the Bekenstein-Hawking entropy $A/(4G)$ of the black hole.

Another potential problem is the negative flux predicted near the peak value of the von Neumann entropy $S$, which would cause the black hole briefly to gain mass instead of losing it monotonically.  However, using the numerical results for Hawking emission of the known massless particles in 4-dimensions leads to a very tiny increase in the mass, less than 0.05\% of the energy of a single quantum of the energy of the Hawking temperature of the black hole at that time.  One can further argue that this tiny effect would be totally swamped by the quantum uncertainty of when in each semiclassical spacetime the von Neumann entropy would reach its peak.

A third potential problem is the fact that extrapolating the expected behavior of the von Neumann entropy of a large black hole down into the final stages in which the black hole gets near the Planck mass leads also to a negative flux of energy.  However, unlike the negative flux at near the peak in $S$, which cannot be avoided in the Bianchi-Smerlak formula, one can modify the assumed behavior of the von Neumann entropy near the final evaporation so that the Bianchi-Smerlak formula continues to give a positive and monotonically rising flux at the end of the evaporation.

The authors are grateful to the Natural Sciences and Engineering Research Council of Canada for financial support.  DNP appreciates discussions with Eugenio Bianchi at Peyresq 20 in Peyresq, France, hosted by OLAM, association pour la recherche fondamentale, Brussels, and also discussions with Matteo Smerlak, hosted by the Perimeter Institute in Waterloo, Ontario, Canada.

\end{document}